\documentclass[10pt, conference, letterpaper]{IEEEtran}

\IEEEoverridecommandlockouts
\usepackage{amsmath,amssymb,amsfonts}
\usepackage{bm}
\usepackage{algorithmic}
\usepackage{cite}
\usepackage{color}
\usepackage{dblfloatfix}
\usepackage{fancyhdr}
\usepackage[flushleft]{threeparttable}
\usepackage{graphicx}
\usepackage{scalerel}
\usepackage{listings}
\usepackage{tabu}
\usepackage{tabularx,multirow,multicol,booktabs,colortbl}
\usepackage{textcomp}
\usepackage{tikz}
\usepackage{xcolor}
\usepackage[acronym]{glossaries}
\usepackage[hyphens]{url}

\usepackage[ruled,linesnumbered]{algorithm2e}
\usepackage{color,soul}

\usepackage{pgfplots}
\usepackage{subcaption}
\usepackage{balance}

\usepackage{pgfplotstable}
\usetikzlibrary[patterns]

\usetikzlibrary[patterns]

\pgfplotsset{
    General/.style={
        width=\linewidth,
        xtick pos=left,
        xtick align=outside,
        ytick pos=left,
        ytick align=outside,
        ymajorgrids=true,
        grid style=dashed,
        legend cell align={left},
        label style={font=\footnotesize},
        legend style={font=\footnotesize},
        ticklabel style={
            font=\footnotesize,
            /pgf/number format/fixed
        },
        enlarge x limits=0.08,
        xtick=data,
        xticklabel style={rotate=45, anchor=east},
    },
    BarConfig/.style={
        General,
        ticklabel style={
            /pgf/number format/fixed zerofill,
            /pgf/number format/precision=1
        }
    },
    BarPercentConfig/.style={
        General,
        yticklabel={\pgfmathprintnumber\tick\%},
        ticklabel style={
            /pgf/number format/precision=0,
            /pgf/number format/fixed zerofill,
        }
    },
    GenPlot/.style={
        height=6cm
    },
    UsePDF/.style={
        height=8cm
    },
}

\definecolor{RWTHBlau100}{RGB}{0,84,159}	
\definecolor{RWTHBlau75}{RGB}{64,127,183}
\definecolor{RWTHBlau50}{RGB}{142,186,229}	
\definecolor{RWTHBlau25}{RGB}{199,221,242}
\definecolor{RWTHBlau10}{RGB}{232,241,250}

\definecolor{RWTHSchwarz100}{RGB}{0,0,0}
\definecolor{RWTHSchwarz75}{RGB}{100,101,103}
\definecolor{RWTHSchwarz50}{RGB}{156,158,159}
\definecolor{RWTHSchwarz25}{RGB}{207,209,210}
\definecolor{RWTHSchwarz10}{RGB}{236,237,237}

\definecolor{RWTHMagenta100}{RGB}{227,0,102}
\definecolor{RWTHMagenta75}{RGB}{233,96,136}
\definecolor{RWTHMagenta50}{RGB}{241,158,177}
\definecolor{RWTHMagenta25}{RGB}{249,210,218}
\definecolor{RWTHMagenta10}{RGB}{253,238,240}

\definecolor{RWTHGelb100}{RGB}{255,237,0}
\definecolor{RWTHGelb75}{RGB}{255,240,85}
\definecolor{RWTHGelb50}{RGB}{255,245,155}
\definecolor{RWTHGelb25}{RGB}{255,250,209}
\definecolor{RWTHGelb10}{RGB}{255,253,238}

\definecolor{RWTHPetrol100}{RGB}{0,97,101}
\definecolor{RWTHPetrol75}{RGB}{45,127,131}
\definecolor{RWTHPetrol50}{RGB}{125,164,167}
\definecolor{RWTHPetrol25}{RGB}{191,208,209}
\definecolor{RWTHPetrol10}{RGB}{230,236,236}

\definecolor{RWTHTuerkis100}{RGB}{0,152,161}
\definecolor{RWTHTuerkis75}{RGB}{0,177,183}
\definecolor{RWTHTuerkis50}{RGB}{137,204,207}
\definecolor{RWTHTuerkis25}{RGB}{202,231,231}
\definecolor{RWTHTuerkis10}{RGB}{235,246,246}

\definecolor{RWTHGruen100}{RGB}{87,171,39}
\definecolor{RWTHGruen75}{RGB}{141,192,96}
\definecolor{RWTHGruen50}{RGB}{184,214,152}
\definecolor{RWTHGruen25}{RGB}{221,235,206}
\definecolor{RWTHGruen10}{RGB}{242,247,236}

\definecolor{RWTHMaigruen100}{RGB}{189,205,0}
\definecolor{RWTHMaigruen75}{RGB}{208,217,92}
\definecolor{RWTHMaigruen50}{RGB}{224,230,154}
\definecolor{RWTHMaigruen25}{RGB}{240,243,208}
\definecolor{RWTHMaigruen10}{RGB}{249,250,237}

\definecolor{RWTHOrange100}{RGB}{246,168,0}
\definecolor{RWTHOrange75}{RGB}{250,190,80 }
\definecolor{RWTHOrange50}{RGB}{253,212,143}
\definecolor{RWTHOrange25}{RGB}{254,234,201}
\definecolor{RWTHOrange10}{RGB}{255,247,234}

\definecolor{RWTHRot100}{RGB}{204,7,30}
\definecolor{RWTHRot75}{RGB}{216,92,65}
\definecolor{RWTHRot50}{RGB}{230,150,121}
\definecolor{RWTHRot25}{RGB}{243,205,187}
\definecolor{RWTHRot10}{RGB}{250,235,227}

\definecolor{RWTHBordeaux100}{RGB}{161,16,53}
\definecolor{RWTHBordeaux75}{RGB}{182,82,86}
\definecolor{RWTHBordeaux50}{RGB}{205,139,135}
\definecolor{RWTHBordeaux25}{RGB}{229,197,192}
\definecolor{RWTHBordeaux10}{RGB}{245,232,229}

\definecolor{RWTHViolett100}{RGB}{97,33,88}
\definecolor{RWTHViolett75}{RGB}{131,78,117}
\definecolor{RWTHViolett50}{RGB}{168,133,158}
\definecolor{RWTHViolett25}{RGB}{210,192,205}
\definecolor{RWTHViolett10}{RGB}{237,229,234}

\definecolor{RWTHLila100}{RGB}{122,111,172}
\definecolor{RWTHLila75}{RGB}{155,145,193}
\definecolor{RWTHLila50}{RGB}{188,181,215}
\definecolor{RWTHLila25}{RGB}{222,218,235}
\definecolor{RWTHLila10}{RGB}{242,240,247}
\usetikzlibrary{svg.path}

\definecolor{orcidlogocol}{HTML}{A6CE39}
\tikzset{
  orcidlogo/.pic={
    \fill[orcidlogocol] svg{M256,128c0,70.7-57.3,128-128,128C57.3,256,0,198.7,0,128C0,57.3,57.3,0,128,0C198.7,0,256,57.3,256,128z};
    \fill[white] svg{M86.3,186.2H70.9V79.1h15.4v48.4V186.2z}
                 svg{M108.9,79.1h41.6c39.6,0,57,28.3,57,53.6c0,27.5-21.5,53.6-56.8,53.6h-41.8V79.1z M124.3,172.4h24.5c34.9,0,42.9-26.5,42.9-39.7c0-21.5-13.7-39.7-43.7-39.7h-23.7V172.4z}
                 svg{M88.7,56.8c0,5.5-4.5,10.1-10.1,10.1c-5.6,0-10.1-4.6-10.1-10.1c0-5.6,4.5-10.1,10.1-10.1C84.2,46.7,88.7,51.3,88.7,56.8z};
  }
}

\newcommand\orcidicon[1]{\href{https://orcid.org/#1}{\mbox{\scalerel*{
\begin{tikzpicture}[yscale=-1,transform shape]
\pic{orcidlogo};
\end{tikzpicture}
}{|}}}}

\usepackage{hyperref}

\def\BibTeX{{\rm B\kern-.05em{\sc i\kern-.025em b}\kern-.08em
    T\kern-.1667em\lower.7ex\hbox{E}\kern-.125emX}}

\def\finalpaper{1} 

\title{Automatic Microarchitecture-Aware Custom Instruction Design for RISC-V Processors}

\if\finalpaper1
\author{\IEEEauthorblockN{
     Evgenii Rezunov,
     Niko Zurstraßen,
     Lennart M. Reimann,
     Rainer Leupers
     }
     \\
     \IEEEauthorblockA{
      \textit{RWTH Aachen University, Institute for Communication Technologies and Embedded Systems} \\
     }
}
\else
\author{
  \IEEEauthorblockN{Authors are removed for submission version}
  \\
  \IEEEauthorblockA{Affiliations are removed for submission version}
}
\fi

\usepackage[pages=some]{background}
\begin{document}

\newacronym[plural=ASICs,firstplural=Application-Specific Integrated Circuits (ASICs)]{asic}{ASIC}{Application-Specific Integrated Circuit}
\newacronym[plural=ASIPs,firstplural=Application-Specific Instruction Set Processors (ASIPs)]{asip}{ASIP}{Application-Specific Instruction Set Processor}
\newacronym[plural=BBs,firstplural=Basic Blocks (BBs)]{bb}{BB}{Basic Block}
\newacronym[plural=CPUs,firstplural=Central Processing Units (CPUs)]{cpu}{CPU}{Central Processing Unit}
\newacronym[plural=Data-Flow Graphs (DFGs)]{dfg}{DFG}{Data-Flow Graph}
\newacronym[plural=Directed Acycles Graphs (DAGs)]{dag}{DAG}{Directed Acyclic Graph}
\newacronym[plural=FSSs,firstplural=Full-System Simulators (FSSs)]{fss}{FSS}{Full-System Simulator}
\newacronym[plural=FUs,firstplural=Functional Units (FUs)]{fu}{FU}{Functional Unit}
\newacronym[plural=GPPs,firstplural=General Purpose Processors (GPPs)]{gpp}{GPP}{General Purpose Processor}
\newacronym[plural=ISAs,firstplural=Instruction Set Architectures (ISAs)]{isa}{ISA}{Instruction Set Architecture}
\newacronym[plural=ISEs,firstplural=Instruction Set Extensions (ISEs)]{ise}{ISE}{Instruction Set Extension}
\newacronym[plural=ISSs,firstplural=Instruction Set Simulators (ISSs)]{iss}{ISS}{Instruction Set Simulator}
\newacronym[plural=MPSoCs,firstplural=Multiprocessor Systems on A Chip (MPSoCs)]{mpsoc}{MPSoC}{Multiprocessor System on A Chip}

\newacronym{eda}{EDA}{Electronic Design Automation}
\newacronym{cid}{CIDRE}{Custom Instruction Designer for RISC-V Extensions}
\newacronym{io}{I/O}{Input/Output}
\newacronym{ilp}{ILP}{Integer Linear Programming}
\newacronym{ip}{IP}{Intellectual Property}
\newacronym{ir}{IR}{Intermediate Representation}
\newacronym{miso}{MISO}{Multiple Input Single Output}
\newacronym{mac}{MAC}{Multiply-Accumulate}
\newacronym{rtl}{RTL}{Register-transfer Level}
\newacronym{simd}{SIMD}{Single Instruction Multiple Data}
\newacronym{cpi}{CPI}{Cycles Per Instruction}
\newacronym{wns}{WNS}{Worst Negative Slack}

\definecolor{funcbrown}{RGB}{116,83,31}
\definecolor{purplekey}{RGB}{143,6,196}
\definecolor{typeturquoise}{RGB}{43,145,175}
\definecolor{purpledefine}{RGB}{138,27,255}
\definecolor{dkgreen}{rgb}{0,0.6,0}
\definecolor{gray}{rgb}{0.5,0.5,0.5}
\definecolor{mauve}{rgb}{0.58,0,0.82}
\definecolor{weborange}{RGB}{255,165,0}
\definecolor{rwth_darkblue}{RGB}{0, 84, 156}
\definecolor{rwth_blue}{RGB}{64, 127, 183}
\definecolor{rwth_lightblue}{RGB}{142, 186, 229}

\lstset{frame=tb,
  language=C++,
  linewidth=8.85cm,
  basicstyle={\scriptsize\ttfamily}, 
  numbers=left,
  numberstyle=\color{gray},
  keywordstyle=\color{blue},
  commentstyle=\color{dkgreen},
  stringstyle=\color{mauve},
  emph={switch,case,break},emphstyle={\color{purplekey}},
  breaklines=true,
  tabsize=2,
  xleftmargin=6.0ex,
  classoffset=4,
  morekeywords={constexpr},keywordstyle=\color{blue},
  classoffset=3,
  morekeywords={likely,QEMU_NO_HARDFLOAT,unlikely,INFINITY},keywordstyle=\color{purpledefine},
  classoffset=2,
  morekeywords={f32, i32, RoundingMode, u32},keywordstyle=\color{typeturquoise},
  classoffset=1,
  morekeywords={},keywordstyle=\color{funcbrown},
  classoffset=0
}

\newcommand\copyrighttext{%
  \footnotesize \textcopyright \the\year{} IEEE. Personal use of this material is permitted. Permission from IEEE must be obtained for all other uses, including reprinting/republishing this material for advertising or promotional purposes, collecting new collected works for resale or redistribution to servers or lists, or reuse of any copyrighted component of this work in other works.}

\newcommand\copyrightnotice{%
    \backgroundsetup{opacity=1, scale=1, angle=0, contents={
            \color{black}%
            \begin{tikzpicture}[remember picture,overlay]%
                \node[anchor=south,yshift=10pt] at (current page.south) {\fbox{\parbox{\dimexpr0.75\textwidth-\fboxsep-\fboxrule\relax}{\copyrighttext}}};
                \node[anchor=north,yshift=-10pt,text=gray] at (current page.north) (preprint) { PREPRINT - Accepted for publication at the 2025 IEEE/ACM International Conference On Computer-Aided Design (ICCAD)};
                \node[anchor=north,yshift=-5pt,text=gray] at (preprint.south) { DOI: 10.1109/ICCAD66269.2025.11240781};
            \end{tikzpicture}%
        }%
    }%
    \BgThispage%
}

\maketitle
\copyrightnotice

\bstctlcite{IEEEexample:BSTcontrol} 

\begin{abstract}
  An \gls{asip} is a specialized microprocessor that provides a trade-off between the programmability of a \gls{gpp} and the performance and energy-efficiency of dedicated hardware accelerators.
  \glspl{asip} are often derived from off-the-shelf \glspl{gpp} extended by custom instructions tailored towards a specific software workload.
  One of the most important challenges of designing an \gls{asip} is to find said custom instructions that help to increase performance without being too costly in terms of area and power consumption.
  To date, solving this challenge is relatively labor-intensive and typically performed manually.

  Addressing the lack of automation, we present \emph{\gls{cid}}, a front-to-back tool for \gls{asip} design.
  \gls{cid} automatically analyzes hot spots in RISC-V applications and generates custom instruction suggestions with a corresponding nML description.
  The nML description can be used with other electronic design automation tools to accurately assess the cost and benefits of the found suggestions.
  In a RISC-V benchmark study, we were able to accelerate embedded benchmarks from Embench and MiBench by up to 2.47x with less than 24\% area increase.
  The entire process was conducted completely automatically.
\end{abstract}

\begin{IEEEkeywords}
  Custom Instructions, Application Specific Instruction Set Processors, Electronic Design Automation, RISC-V
  \end{IEEEkeywords}
\vspace{-0.2cm}

\glsresetall
\section{Introduction}
\label{section:introduction}
When maximum performance and energy efficiency are required for a specific task, designers often resort to so-called \glspl{asic}, which implement hardwired data paths and aggressive hardware optimizations.
However, their lack of flexibility makes them unsuitable for applications that demand programmability.
As a compromise, \glspl{asip} provide a middle ground between \glspl{asic} and \glspl{gpp} by extending a baseline \gls{gpp} with custom instructions tailored to specific workloads.
These customizations are designed to accelerate performance-critical operations, while the base \gls{isa} handles the remaining functionality of the processor.
Custom instructions are typically implemented in hardware as \glspl{fu} integrated into the processor's pipeline~\cite{ienne2006customizable}.

As \glspl{asip} require a base processor, computer architects have long been reliant on proprietary \gls{ip} and \glspl{isa}, such as Cadence Tensilica Xtensa~\cite{xtensa}, and Synopsys ARC~\cite{synopsysarc}.
With the emergence of the open RISC\nobreakdash-V \gls{isa}~\cite{riscvOGPaper} and the open software and hardware ecosystem that developed around it,
computer architects and researchers were able to develop \glspl{asip} that were not based on the products of a single company.
This has uniquely changed the market and led to a renaissance of \gls{asip} research.

Processor customization remains a challenging task that requires a deep understanding of both the target application and the underlying architecture.
The increasing demand for application-specific processors and the labor-intensive nature of customization have driven extensive research into automating various steps of the \gls{asip} design process over the past decades~\cite{hussein2024automating}.
However, many existing approaches focus primarily on algorithmic aspects and overlook microarchitectural considerations.
A comprehensive solution that is integrated into an \gls{isa} ecosystem like RISC\nobreakdash-V and accounts for microarchitectural constraints and trade-offs is still lacking.

In this work, we present \emph{\gls{cid}}: an automatic front-to-back tool for \gls{asip} design.
\gls{cid} analyzes hot spots in RISC\nobreakdash-V applications and generates custom instruction suggestions with a corresponding nML~\cite{van2008nml} processor model.
The nML description can be used to accurately assess the cost and performance gain from the found suggestions.
In a RISC\nobreakdash-V benchmark study, we were able to accelerate embedded benchmarks from Embench~\cite{embench} and MiBench~\cite{guthaus2001mibench} by up to 2.47x with less than 24\% area increase.
Using predefined constraint setups, the whole process from an application executable to a synthesized processor can be conducted without any manual labor.

\section{Background \& Related Work}
\label{section:background}
The process of finding custom instructions consists of three main phases.
The first phase, \emph{instruction enumeration}, identifies all potential candidates in a given application.
To eliminate duplicates, the second phase, \emph{isomorphism detection}, clusters the candidates in equivalence classes.
Finally, the \emph{instruction selection} phase selects the most beneficial instructions from these classes.
The following section introduces key terminology, discusses each phase in detail, and provides an overview of how state-of-the-art methods address these challenges.

\subsection{Data-Flow Graphs}
\label{subsection:data-flow-graphs}
A software application typically consists of multiple basic blocks -- sequences of instructions with one entry and one exit point.
Each basic block can be represented as a \gls{dfg}. A \gls{dfg} is a directed acyclic graph $G(V, E)$, where the vertex set $V=\{v_1,..., v_n\}$ denotes operations,
and the edge set $E=\{e_1,..., e_m\} \in (V\times V)$ represents the data dependencies between them.
Inputs and outputs of the \gls{dfg} -- either vertices from other basic blocks or immediate operands -- are represented as the vertex set $V^{+}$, with their corresponding edges $E^{+}$.
Together with $G$, these form the extended graph $G^{+}(V \cup V^{+}, E \cup E^{+})$ \cite{xiao2012exact}.

\begin{figure}[t]
  \centering
  \includegraphics[width=0.47\textwidth]{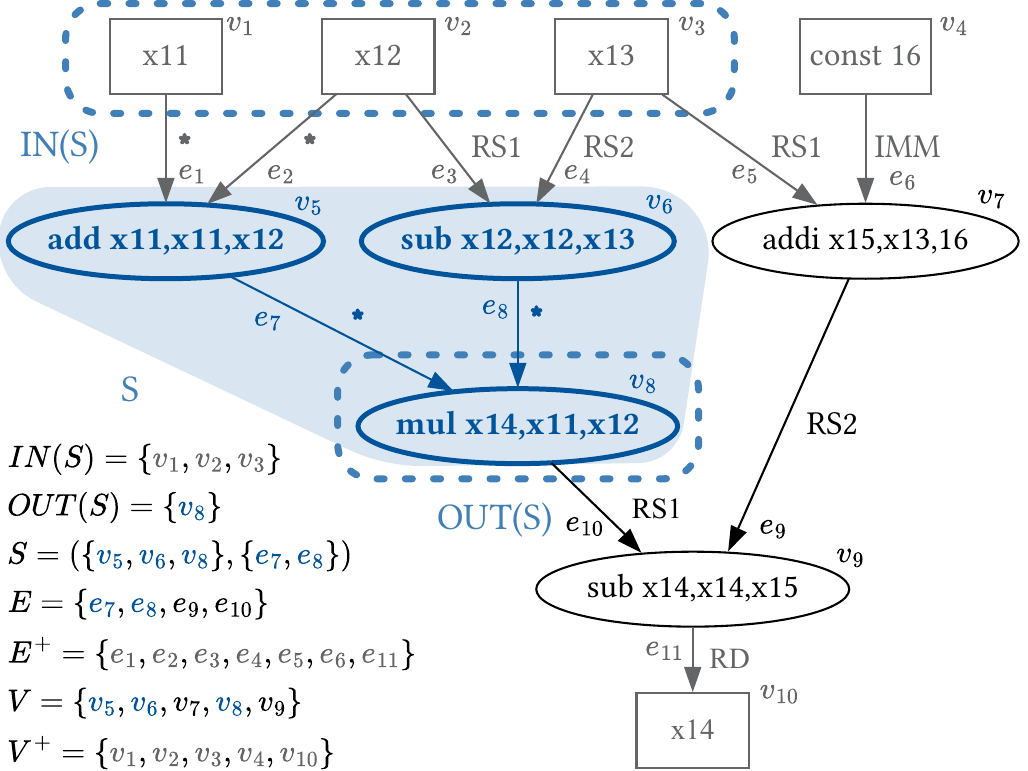}
  \caption{Example of a \gls{dfg}.}
  \label{fig:example-pattern}
  \vspace{-0.3cm}
\end{figure}

Fig.~\ref{fig:example-pattern} provides an example of a \gls{dfg} used in this work.
Since it targets RISC\nobreakdash-V, the vertices $V$ represent RISC\nobreakdash-V instructions.
The external vertices $V^{+}$ model either registers (e.g, x0-x31 or f0-f31) or immediate values.
To preserve operand semantics, the edges $E$ and $E^{+}$ carry information about operand positions, since operations are generally not commutative.
Depending on the type, RISC\nobreakdash-V instructions may use up to three source registers (RS1, RS2, and RS3), one immediate input (IMM), and one destination register (RD).
Commutative operations (e.g., ADD, AND, MUL) are marked with an asterisk (*) on the corresponding edges.

A custom instruction is defined as an induced subgraph $\mbox{$S \subseteq G$}$.
We use $IN(S)$ to denote inputs -- external vertices with edges entering $S$ -- and $OUT(S)$ for the outputs -- vertices inside $S$ with edges exiting the subgraph.
Fig.~\ref{fig:example-pattern} also illustrates an example subgraph $S$ representing a custom instruction.

\subsection{Instruction Enumeration}
\label{subsection:instruction-enumeration}
Instruction enumeration is the process of identifying valid instruction clusters within the application's \glspl{dfg}.
Since each vertex of $G$ can either be included or excluded from $S$, the maximum number of possible patterns is $2^{|V|}$.
However, the number can be significantly reduced by imposing constraints.
In fact, as shown in \cite{chen2007fast} and \cite{bonzini2007polynomial}, constrained valid patterns $S=\{S_1,...,S_n\}$ are polynomial in the number of vertices of the \gls{dfg}.
The constraints usually include:

\emph{\gls{io} constraints}:
An important factor limiting the number of inputs $IN(S)$ and outputs $OUT(S)$ for a custom instruction is the number of read and write ports in the register file.
Another constraint is the number of bits available in the instruction encoding.
For instance, the base RISC\nobreakdash-V \acrshort{isa} uses 32-bit instructions that must encode the opcode, source and destination registers or immediates, and additional instruction-specific bits. The maximum number of input and output operands supported by a custom instruction is denoted by $IN_{\text{max}}$ and $OUT_{\text{max}}$, respectively.

\begin{figure}[t]
  \centering
  \includegraphics[width=0.45\textwidth]{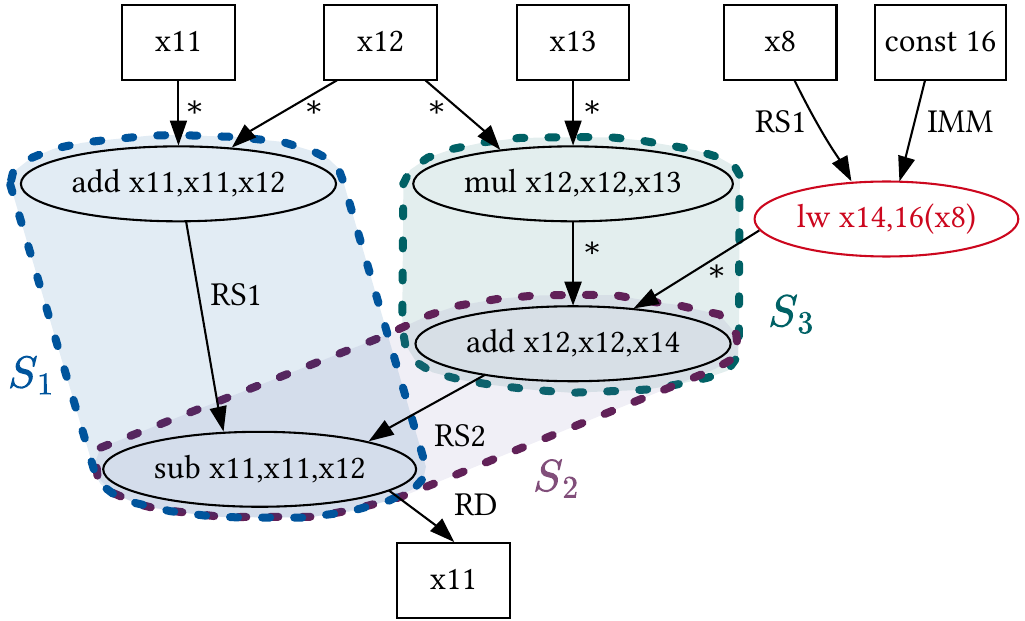}
  \caption{Example of instruction enumeration. Constraints ($IN_{\text{max}} = 3$, $OUT_{\text{max}} = 1$, $lw \in F$) reduce the search space from $2^5=32$ to $3$ patterns (one-vertex patterns excluded).}
  \label{fig:instruction-enumeration}
  \vspace{-0.3cm}
\end{figure}

\emph{Convexity constraint}: For an instruction cluster to be executable as a single instruction, it must be schedulable.
In graph theory terms, this means that the corresponding subgraph $S$ must be convex -- there must be no path between two vertices $v, u \in S$ that passes through any vertex $w \notin S$~\cite{pozzi2006exact}.
For instance, in Fig.~\ref{fig:instruction-enumeration}, a disconnected subgraph containing the SUB and MUL operations is not convex and therefore invalid.

\emph{Forbidden instructions}: Depending on the architecture of the \gls{fu}, certain instructions may be excluded from custom instruction candidates.
For example, if the \gls{fu} lacks memory ports, corresponding instructions must be omitted (e.g., the LW operation in Fig.~\ref{fig:instruction-enumeration}).

The problem of instruction enumeration has been widely studied, with various works proposing different formulations based on specific algorithms and performance objectives.
For example, \cite{cong_application-specific_2004} and \cite{galuzzi2007linear} focus on enumerating patterns where the output constraint $OUT_{\text{max}}$ is explicitly set to 1 in order to enhance algorithm performance.
Other studies, such as \cite{atasu2008fast}, \cite{atasu2012fish}, and \cite{giaquinta_maximum_2015}, concentrate on enumerating maximal subgraphs -- subgraphs that cannot be further expanded by including additional vertices from the vertex set $V$.
These works argue that the speedup achieved through custom instructions is monotonic with the size of the subgraph, and thus, larger subgraphs tend to offer better performance improvements~\cite{verma2007rethinking}.
Finally, works such as \cite{pozzi2006exact} and \cite{xiao2012exact} focus on exhaustive algorithms that enumerate all possible subgraphs within the constraints. While these algorithms are computationally expensive, they guarantee that no valid subgraph is overlooked.

In summary, instruction enumeration is largely considered solved, with existing algorithms providing effective solutions for detecting valid patterns. However, their performance remains constrained by the problem's complexity.

\begin{figure}[t]
  \centering
  \includegraphics[width=0.43\textwidth]{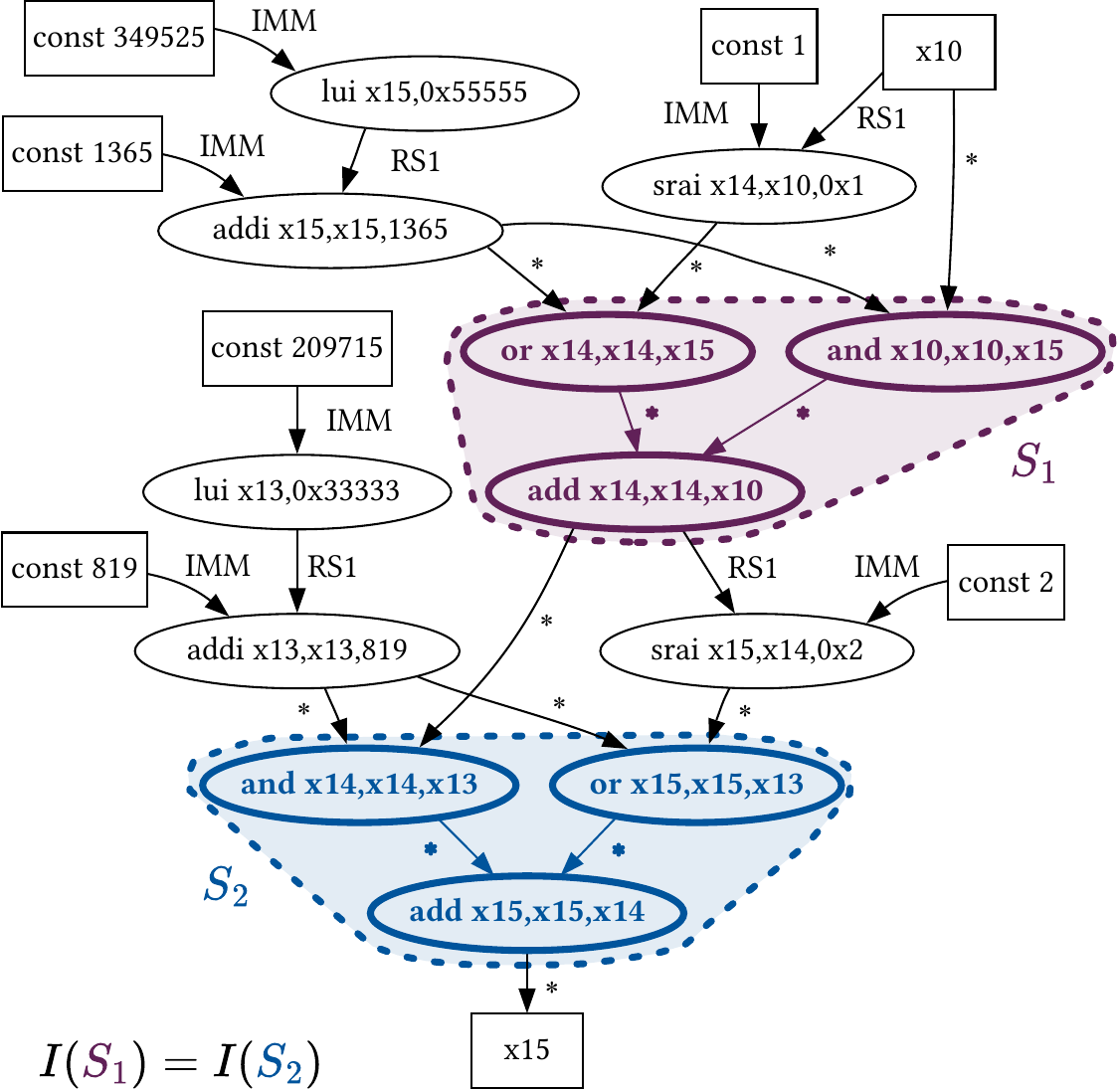}
  \caption{Example of two isomorphic subgraphs in a \gls{dfg}. The basic block was extracted from MiBench's \textit{bitcnts} benchmark.}
  \label{fig:isomorphism-example}
  \vspace{-0.3cm}
\end{figure}

\subsection{Isomorphism Detection}
\label{subsection:isomorphism-detection}
Once all valid subgraphs are identified, the next step is to detect equivalent ones, allowing for hardware reuse. Equivalent patterns correspond to recurring custom instructions that may appear within a single basic block, an application, or even across multiple applications. Fig.~\ref{fig:isomorphism-example} illustrates two equivalent patterns in a basic block.

In terms of graph theory, detection of equivalent instruction patterns corresponds to the subgraph isomorphism problem.
Two (sub-) graphs $G(V, E)$ and $G'(V', E')$ are isomorphic if there exists a bijection $f: V \rightarrow V'$ such that for every edge $(v_i, v_j) \in E$, there exists an edge $(f(v_i), f(v_j)) \in E'$~\cite{groheGraphIsomorphismProblem}.
In the context of instruction customization, this definition has to be expanded by two criteria:
\begin{itemize}
    \item The vertices of the graph are mapped onto vertices that implement the same operation.
    This can be seen in Fig.~\ref{fig:isomorphism-example} (AND $\rightarrow$ AND, OR $\rightarrow$ OR, ADD $\rightarrow$ ADD).
    \item Edges in the graph are mapped to corresponding edges with identical operand positions. For example, if the edge $(v_i, v_j)$ is labeled as RS1, then $(f(v_i), f(v_j))$ must also be labeled as RS1. Note that the ADD operation in Fig.~\ref{fig:isomorphism-example} is commutative, so the operands can be swapped without affecting the isomorphism check.
\end{itemize}
All equivalent subgraphs are grouped into isomorphism classes, denoted $U$. The notation $I(S) = U$ indicates that subgraph $S$ belongs to equivalence class $U$.

Graph isomorphism is a problem that is not yet proven to be solvable in polynomial time~\cite{groheGraphIsomorphismProblem}.
Algorithms like VF2~\cite{cordella2004sub} are commonly used to detect isomorphic subgraphs and have been applied to custom instruction generation, for example by Bonzini and Pozzi~\cite{bonzini2008recurrence}.
An alternative approach is subgraph canonization, as proposed by Ahn and Choi~\cite{ahn2012isomorphism}.
The \emph{canonical form} of a (sub-)graph $S$, denoted $CF(S)$, is a unique representation such that $CF(S_1) = CF(S_2)$ if and only if $S_1$ and $S_2$ are isomorphic.

\begin{figure}[t]
  \centering
  \includegraphics[width=0.45\textwidth]{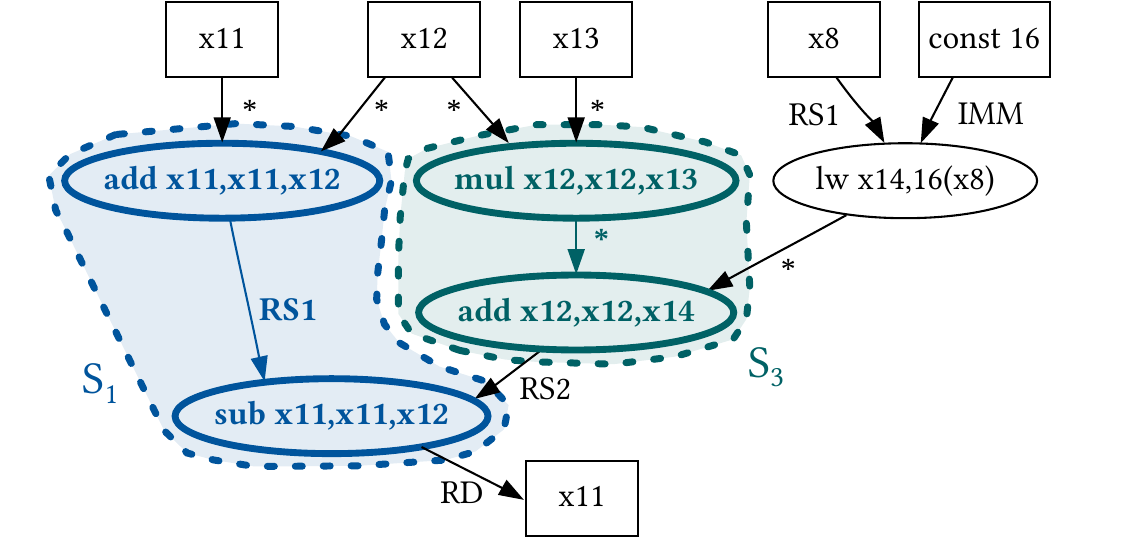}
  \caption{Instruction selection for subgraphs from Fig.~\ref{fig:instruction-enumeration}.}
  \label{fig:instruction-selection-example}
  \vspace{-0.3cm}
\end{figure}

\subsection{Instruction Selection}
\label{subsection:instruction-selection}
The final step in the customization process is selecting the most beneficial instructions, constrained by a predefined maximum number of equivalence classes, $U_{\text{max}}$.
For this purpose, a merit function $M(S)$ evaluates the performance gain of implementing a pattern $S$ as a single custom instruction.
If simulation-based profiling data with basic block execution counts is available, the merit for each subgraph is scaled accordingly.
This way, optimizations in performance-critical regions are prioritized over less frequently executed code.

The merit function is commonly defined as the difference in execution time before and after customization:
\begin{equation*}
  M(S) = L_{\text{SW}}(S) - L_{\text{HW}}(S)
\end{equation*}
where $L_{\text{SW}}(S)$ is the sum of individual operation latencies (in processor cycles) when the subgraph is executed in software, and $L_{\text{HW}}(S)$ is the latency of the corresponding custom instruction implemented in hardware~\cite{ahn2012isomorphism, atasu2012fish, wang2023reinforcement}.

A simple estimation assumes that all custom instructions execute in a single cycle~\cite{clark2006scalable}, but
more precise methods consider the critical path of the subgraph.
Works such as~\cite{clark2003processor, atasu2012fish, atasu2007optimizing, wang2023reinforcement, wang2015selecting, pozzi2006exact, atasu2008chips}
determine the hardware latency of each vertex by synthesizing the corresponding operation and normalizing the resulting delay to a reference, such as a 32-bit integer addition or a multiply-accumulate operation. The total latency is then calculated as the sum of the normalized values along the critical path.
Finally, some works, such as~\cite{atasu2008fast}, perform hardware synthesis of the entire custom instruction, improving estimation accuracy at the cost of increased runtime.
\begin{figure*}[t]
  \centering
  \includegraphics[width=0.90\textwidth]{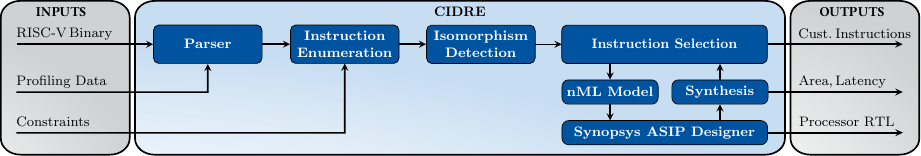}
  \caption{Overview of the framework. Profiling data is obtained from an instruction-accurate simulator.}
  \label{fig:framework-flow}
  \vspace{-0.3cm}
\end{figure*}

Graph covering is an essential part of the instruction selection step, as the subgraphs are not allowed to conflict with each other. Two constraints are imposed on the candidates~\cite{wang2023reinforcement}:
\begin{itemize}
    \item \emph{Non-overlapping constraint}: The selected custom instructions must not share nodes in the \gls{dfg}.
    Examples of subgraphs violating the non-overlapping constraint are shown in Fig.~\ref{fig:instruction-enumeration} ($S_1$, $S_2$ and $S_2$, $S_3$).
    \item \emph{Acyclicity constraint}: Cycles can occur between the subgraph candidates if they include disconnected components.
    If two subgraphs provide data for each other, it leads to a deadlock situation.
\end{itemize}

Essentially, the task of the isomorphism-aware instruction selection can be described as the problem of selecting a set of enumerated subgraphs corresponding to at most $U_{\text{max}}$ equivalence classes that maximize the total merit function while satisfying the non-overlapping and acyclicity constraints.

Fig.~\ref{fig:instruction-selection-example} illustrates an optimal instruction selection for the instructions from Fig.~\ref{fig:instruction-enumeration}, assuming all instruction clusters execute in a single cycle.
Note that selecting $S_2$ would not yield an optimal covering, as it would block the selection of other instructions due to the non-overlapping constraint.

Optimal code selection on directed acyclic graphs is a well-known NP-complete problem~\cite{aho1976code}.
The number of feasible solutions grows exponentially with the number of subgraphs~\cite{wang2015selecting}.
Due to this high complexity, most approaches represent a trade-off between optimality and efficiency~\cite{hussein2024automating}.

Several approaches~\cite{atasu2012fish, atasu2005integer, karuri_application_2011} use integer linear programming to solve the problem.
However, due to the problem's high complexity, this approach is typically not used to select an optimal solution from exhaustively enumerated subgraphs.

Heuristic algorithms, on the other hand, provide better scalability and efficiency but may not always find the optimal solution.
Many heuristics rely on a greedy approach, where subgraphs are selected iteratively based on their merit function~\cite{ahn2012isomorphism}.
Other methods use meta-heuristic approaches, such as genetic algorithms~\cite{pozzi2006exact, xiao2014automatic}, tabu search~\cite{li2010selecting}, or ant colony optimization~\cite{wang2015selecting}.
Machine learning methods, like reinforcement learning, have also been explored~\cite{wang2023reinforcement}.
Some approaches combine both exact and heuristic algorithms to achieve a good balance between optimality and efficiency~\cite{bonzini2008recurrence}.

\subsection{Shortcomings of Existing Approaches}

While existing approaches to custom instruction generation provide a solid foundation, they face several limitations when applied to real hardware architectures:

\emph{Absence of a reference \gls{isa}}:
Most existing approaches use the compiler-generated \gls{ir} as input to their customization toolchains.
While this avoids additional dependencies in the \glspl{dfg} created by scheduling and register allocation~\cite{clark2003processor}, it presents limitations.
The compiler \gls{ir} still contains operations that require legalization.
For example, in RISC\nobreakdash-V, loading a 32-bit immediate must be broken into a 20-bit upper load and a 12-bit lower addition.
Additionally, the target architecture may impose restrictions on custom operations.
Instruction encoding limits the number of \gls{io} registers, as well as the number and width of immediate values.
Moreover, not all encoding bits are available for custom instructions, since some are reserved.

\emph{Inaccurate merit estimation}:
Existing approaches often overlook key microarchitectural aspects, such as the integration of the instructions in the processor's pipeline and register file accesses.
Furthermore, they do not account for the processor's critical path. If a custom instruction has a shorter critical path than the processor, it cannot execute faster than the clock cycle. Conversely, if its critical path is longer, it may either slow down the processor or require pipelining, introducing synchronization overhead.
Ultimately, we believe that an accurate performance and area estimation requires integrating custom instructions into the processor's microarchitecture.

\emph{Limited scope}:
While many works address individual aspects of custom instruction generation, to the best of our knowledge, no tool offers a complete end-to-end solution.
\section{Methods \& Implementation}
\label{section:methods}

\begin{figure*}[t]
  \centering
  \includegraphics[width=\textwidth]{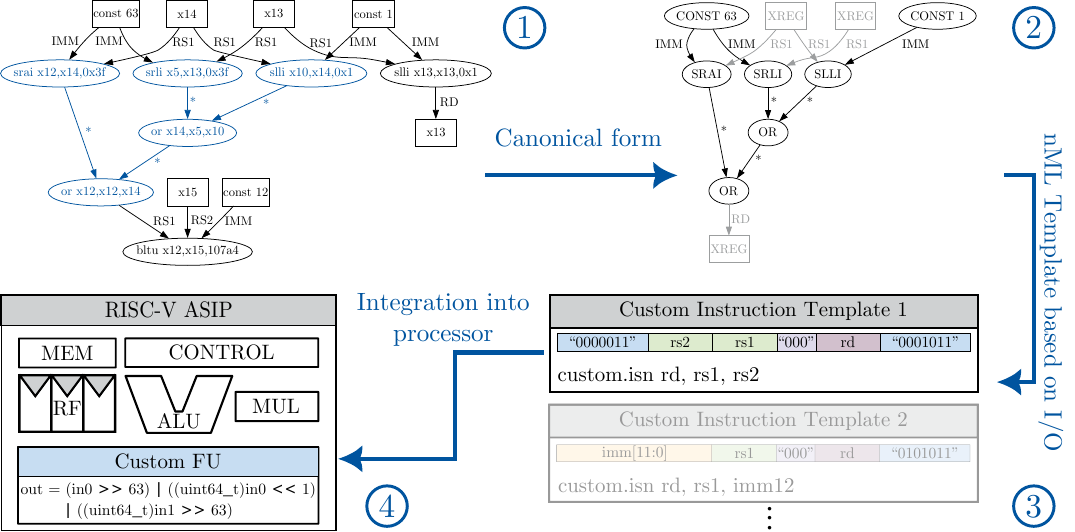}
  \caption{Example of a custom instruction in Embench's \textit{aha-mon64} benchmark. The enumerated subgraph (1) is converted into its canonical form (2). An nML template (3) is selected based on the subgraph's \gls{io} configuration: two register inputs and one register output.  The instruction is integrated into the processor model with corresponding functionality in PDG (4).}
  \label{fig:isn-example}
  \vspace{-0.3cm}
\end{figure*}

In the light of the discussed shortcomings, we propose a framework called \gls{cid} that adheres to following principles. \emph{\gls{isa} awareness} is ensured by working directly with RISC\nobreakdash-V binaries, preserving architectural constraints. \emph{Accurate merit estimation} is achieved by integrating generated instructions into a processor model. This way, microarchitecture details are considered. Finally, \emph{\gls{cid} provides an end-to-end solution that automates the entire process from a RISC\nobreakdash-V binary to a functional \gls{asip} model implementing custom instructions.}

The tool flow is illustrated in Fig.~\ref{fig:framework-flow}.
\gls{cid} takes as input a RISC\nobreakdash-V binary compiled for the baseline processor, along with profiling data containing execution counts for each basic block of the application.
This profiling data can be obtained from instruction-accurate simulators. 
Additionally, \gls{cid} accepts user-defined constraints, specifying the \gls{io} configuration and the maximum number of custom instructions to be generated.

A custom disassembler is implemented to process the RISC\nobreakdash-V binary compiled for the baseline processor.
Subsequently, control flow analysis is performed to detect basic blocks, which are then converted into \glspl{dfg}.

An exhaustive approach is used for instruction enumeration, ensuring all possible subgraphs are considered. This enhances the framework's modularity by imposing no restrictions beyond user-defined constraints, leaving instruction filtering to later stages. To efficiently solve the exhaustive enumeration problem, we adopt the algorithm by Xiao and Casseau~\cite{xiao2012exact}.
The algorithm recursively constructs subgraphs starting from one-node patterns consisting of each valid vertex in the \gls{dfg}. The patterns are then extended by adding further valid vertices in reverse topological order.
To reduce the search space, the algorithm filters out vertices that cannot be added to the subgraph due to the convexity or \gls{io} constraints. Both connected and disconnected patterns are considered, exploiting the potential benefits of parallelism in the custom instruction.

The number of source and destination operands for custom instructions is limited by user-defined \gls{io} constraints.
Immediate values are treated as instruction inputs, and \gls{cid} imposes restrictions on the number and width of those, depending on the \gls{io} configuration and available encoding space.
If a subgraph exceeds the constraints, our tool treats excess immediate values as hardcoded, enabling the selection of larger subgraphs within the given limits.
An example of such a subgraph is shown in Fig.~\ref{fig:isn-example} (1), assuming the \gls{io} restrictions are set to two inputs and one output.

Regarding forbidden operations, \gls{cid} omits load/store instructions due to costly memory access and jump/branch instructions due to their effect on control flow.

After enumeration, the isomorphism detection step groups the subgraphs into equivalence classes using the graph canonization approach by Ahn and Choi~\cite{ahn2012isomorphism}.
Their method employs a branch-and-bound technique to sort the vertices and edges of the subgraph in lexicographically minimal order, ensuring a unique representation for all isomorphic subgraphs.
Compared to algorithms like VF2, this approach has a similar time complexity but requires only $O(n)$ canonical form constructions to group $n$ subgraphs, rather than $O(n^2)$ pairwise comparisons.
Additionally, the canonical form enables efficient storage and retrieval of subgraphs, serving as a key in a hash table to store subgraph occurrences.
Fig.~\ref{fig:isn-example} (2) shows the graphical representation of the canonical form of the subgraph in (1).

Finally, the instruction selection process aims to choose a set of instructions (i.e., isomorphism classes) that maximizes the estimated application speedup while adhering to a user-defined limit $U_{\text{max}}$ on the total number of instructions.

Unlike most approaches that pre-synthesize primitive operations to estimate performance gain, \gls{cid} integrates custom instructions into an nML processor model. This model provides a high-level description of the processor's microarchitecture, including pipeline stages, register file, and supported \gls{isa}. Instruction functionality is described using a C-like language called PDG.
We use Synopsys ASIP Designer~\cite{SynopsysASIPDesigner} to generate \gls{rtl} code for the processor, which is synthesized to assess performance gain and area.
This approach accounts for the instruction's impact on the decoder, register file, pipeline, and critical path.

ASIP Designer provides baseline processor models for customization. For our implementation, we use the 64-bit RISC\nobreakdash-V processor model \textit{trv64p3}, which features a three-stage pipeline and supports the RV64IM \gls{isa}.

\begin{figure}[t]
  \centering
  \includegraphics[width=\columnwidth]{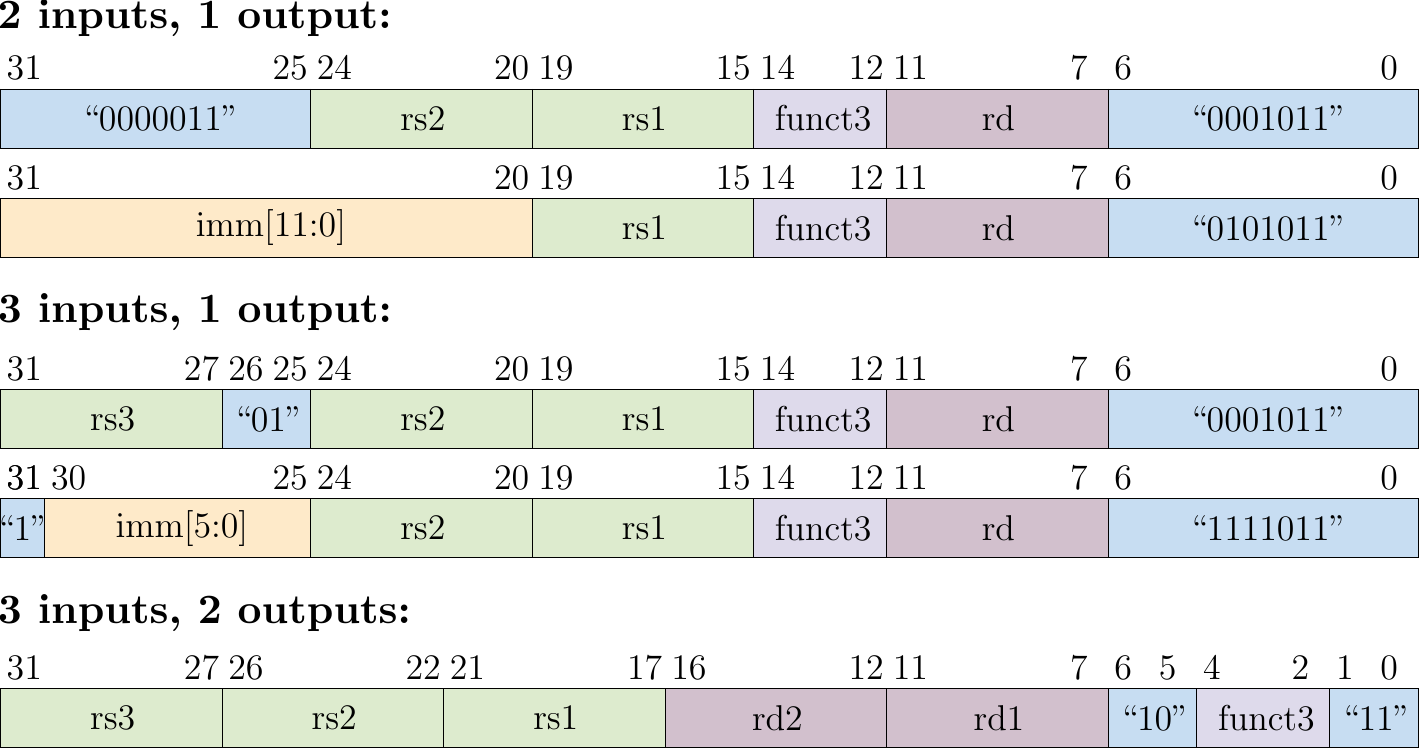}
  \caption{Encodings used for custom instructions.}
  \label{fig:encodings}
  \vspace{-0.3cm}
\end{figure}

Each custom instruction needs to be implemented in nML. Different \gls{io} constraints require distinct instruction encodings, as well as variations in the number of register-file ports and supporting logic. We define five types of instructions, supporting different numbers and types of operands, as demonstrated in Fig.~\ref{fig:encodings}.
Similar to existing RISC\nobreakdash-V instructions, the encodings use 5-bit \textit{rs} and \textit{rd} fields for source and destination registers and a 6- or 12-bit \textit{imm} field for immediate operands. The 3-bit \textit{funct3} field allows for up to eight different instructions for each encoding type.
Following \gls{io} constraints are supported ($IN_{\text{max}}, OUT_{\text{max}}$):

\emph{(2, 1):} Two instruction types are defined corresponding to RISC-V's \textit{R-type} and \textit{I-type} instructions with \textit{CUSTOM\nobreakdash-0} and \textit{CUSTOM\nobreakdash-1} opcodes.

\emph{(3, 1):} The first configuration supports three source registers and one destination register, corresponding to RISC-V's \textit{R4-type} instructions. The second configuration modifies the encoding to include one 6-bit immediate operand. The opcodes used are \textit{CUSTOM\nobreakdash-0} and \textit{CUSTOM\nobreakdash-3}.

\emph{(3, 2):} The final configuration requires 25 bits for the five registers, which exceeds the limitations set by the official RISC\nobreakdash-V guidelines. Since the used baseline processor does not support the floating-point extension, its encoding space is utilized for the \textit{funct3} field.

The five instruction types are dynamically included or excluded from the target processor during selection, as illustrated in Fig.~\ref{fig:isn-example} (3). The PDG primitives describing the instructions' functionality are automatically generated from the instruction's canonical representation, as shown in Fig.~\ref{fig:isn-example} (4).

Similarly to existing approaches, the performance gain of each custom instruction is estimated by comparing the execution time of the corresponding base operations in software with the custom instruction implemented in hardware.
For software execution time \(L_{\text{SW}}\), it is assumed that every operation in the custom instruction subgraph can be executed in one cycle, corresponding to a \gls{cpi} value of 1:
\begin{equation*}
  L_{\text{SW}}(S) = \sum_{v \in S} s(v) = |S|.
\end{equation*}

\(L_{\text{HW}}\) is estimated as one cycle, assuming that each custom instruction takes the same time to execute as other instructions in the processor.
This way, it is ensured no complex synchronization mechanism is needed to integrate the custom instruction into the processor's pipeline, resulting in a more accurate estimation of the performance gain.

\begin{algorithm}[b!]
  \SetFuncSty{textit}
\SetKw{KwContinue}{continue}
\SetKw{KwBreak}{break}

\SetAlgoLined
\KwIn{Candidate isomorphism classes $U_{candidates}$}
\KwOut{Selected isomorphism classes $U_{selected}$}

\BlankLine

\SetKwFunction{FSelection}{instructionSelection}
\SetKwFunction{FExTime}{CalcExTime}
\SetKwFunction{FSynth}{getSynthesisClkPeriod}
\SetKwFunction{FGetBest}{chooseBestInstruction}
\SetKwFunction{FRemove}{removeInstruction}
\SetKwProg{Fn}{Function}{:}{}

\BlankLine
\Fn{\FSelection{$U_{candidates}$}}{

  \While {$|U_{selected}| < U_{max}$}{
    $T_{ex, best} \gets T_{ex, base}$\;
    $U_{current} \gets U_{candidates}$\;
    \Repeat {$T_{clk, new} = T_{clk, base}\  \| \ T_{ex, new} \geq T_{ex, best}$}{
      $u \gets$ \FGetBest{$U_{current}$}\;
      $T_{clk, new} \gets$ \FSynth{$u$}\;
      $T_{ex, new} \gets$ \FExTime{$u, T_{clk, new}$}\;

      \If {$T_{ex, new} \geq T_{ex, base}$}{
        \FRemove{$U_{current}, u$}\;
        \KwContinue\;
      }

      \If {$T_{ex, new} < T_{ex, best}$}{
        $T_{ex, best} \gets T_{ex, new}$\;
        $u_{best} \gets u$\;
        \FRemove{$U_{current}, u$}\;
      }
    }

    $U_{selected} \gets U_{selected} \cup u_{best}$\;
    \FRemove{$U_{candidates}, u_{best}$}\;
  }
}

  \caption{Proposed instruction selection algorithm}
  \label{alg:ch4_synth_selection}
\end{algorithm}

Then, the cycle-based merit of an instruction represented as an isomorphism class \(U\) is calculated as:
\begin{equation*}
  M(U) = \sum_{S \in S_{U}} M(S) \cdot f_{\text{BB, S}} = \sum_{S \in S_{U}} (|S| - 1) \cdot f_{\text{BB, S}}
  \label{eq:ch4:saved_cycles}
\end{equation*}
where \(S_{U} = \{S : I(S) = U\}\) is the set of subgraphs with a feasible cover corresponding to the isomorphism class \(U\), and \(f_{\text{BB, S}}\) is the execution count of the basic block containing \(S\).

To account for the changes in the processor's microarchitecture, the cycle-based merit function is extended by the latency values gained from synthesis.
The execution time of the application is then estimated by multiplying the total execution count of the instructions in the application by the clock period, assuming a \gls{cpi} value of 1:
\begin{equation*}
  T_{\text{execution}} = f_{\text{total}} \cdot T_{\text{clk}}.
  \label{eq:execution_time}
\end{equation*}

A synthesis framework around Synopsys Design Compiler~\cite{SynopsysDC} is utilized to determine the processor's clock period. The selected value corresponds to the highest achievable frequency using the given synthesis configuration.

Since $M(U)$ represents the number of cycles saved by the custom instruction, the estimated execution time of the application on the customized processor is:
\begin{equation*}
  T_{\text{execution, custom}}(U) = (f_{\text{total}} - M(U)) \cdot T_{\text{clk, custom}}
  \label{eq:ch4_custom_execution_time}
\end{equation*}
where $T_{\text{clk, custom}}$ is the clock period of the customized processor. It is greater than or equal to that of the baseline processor. To determine if the custom instruction increases the processor's critical path, the \gls{asip} is synthesized for the same clock period as the baseline processor.
If this timing constraint cannot be met, the next fastest feasible clock period is used.
The instruction yielding the lowest $T_{\text{execution, custom}}$ provides the highest performance gain.

The selection process is shown in Algorithm~\ref{alg:ch4_synth_selection}. At each step, the instruction with the highest merit is selected (Line~6).
The processor model is then synthesized to determine the corresponding clock period and application execution time. If the execution time is worse compared to the baseline processor, the instruction is discarded and the search continues (Lines~10-11).
If it improves performance, it is saved as the new best instruction.
However, if the customized processor's clock period exceeds that of the baseline, the search continues to check if the next best instruction offers a greater speedup~(Line~18). The algorithm terminates once the desired number of instructions is selected.

The selection algorithm follows a greedy heuristic, which can be suboptimal since an instruction chosen early may block better candidates later.
To address this, the two-optimal approach by Bonzini and Pozzi~\cite{bonzini2008recurrence} is used.
Instead of picking an instruction independently of other candidates in the \textit{chooseBestInstruction} function, it optimally selects two candidates and returns the one yielding the highest speedup.
This one-step look-ahead improves the overall selection process.

\newcommand{\plotarea}[1]{
    \pgfplotstableread[col sep=comma]{#1}\synthresults

    \newcommand{\barwidth}{0.035\textwidth}

    \begin{tikzpicture}
        \begin{axis}[
            BarPercentConfig,
            GenPlot,
            height=4cm,
            clip bounding box=upper bound,
            bar width=\barwidth,
            ybar,
            yminorgrids=true,
            ymin=0,
            ytick distance=10,
            xticklabels from table={\synthresults}{benchmark},
            ymax=63,
            transpose legend,
            legend to name=arealegend,
            legend columns=-1,
            legend style={inner sep=5pt,draw=black, fill=white, fill opacity=0.5, text opacity=1},
            ]

            \addplot[bar shift=-\barwidth/2, fill=RWTHBlau100] table [x expr=\coordindex, y=time-reduction] {\synthresults};
            \addlegendentry{Execution time reduction}


            \addplot[bar shift=+\barwidth/2, fill=RWTHBordeaux50] table [x expr=\coordindex, y=area-overhead] {\synthresults};
            \addlegendentry{Area overhead}

        \end{axis}
    \end{tikzpicture}
}

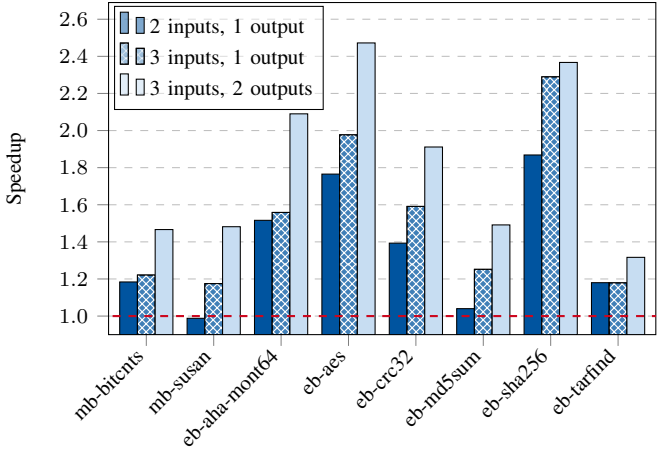
\begin{figure}[t]
  \centering
  \pgfplotstableread[col sep=comma]{tikz/data/synth_io.csv}\synthresults

\newcommand{\barwidth}{0.013\textwidth}

\begin{tikzpicture}
    \begin{axis}[
        BarConfig,
        GenPlot,
        clip bounding box=upper bound,
        bar width=\barwidth,
        ybar,
        yminorgrids=true,
        ymin=0.9,
        ymax=2.69,
        ytick distance=0.2,
        xticklabels from table={\synthresults}{benchmark},
        ylabel={Speedup},
        transpose legend,
        legend style={at={(0.01,0.99)}, anchor=north west, legend columns=1, inner sep=2pt,draw=black, fill=white, fill opacity=0.5, text opacity=1},
        ]

        \addplot[bar shift=-\barwidth, fill=RWTHBlau100] table [x expr=\coordindex, y=2_1] {\synthresults};
        \addlegendentry{2 inputs, 1 output}

        \addplot[bar shift=0cm, fill=RWTHBlau75, postaction={pattern=crosshatch, pattern color=white}] table [x expr=\coordindex, y=3_1] {\synthresults};
        \addlegendentry{3 inputs, 1 output}

        \addplot[bar shift=+\barwidth, fill=RWTHBlau25] table [x expr=\coordindex, y=3_2] {\synthresults};
        \addlegendentry{3 inputs, 2 outputs}

        \draw[RWTHRot100, dashed, thick] (axis cs:-0.5,1) -- (axis cs:10.5,1);

    \end{axis}
\end{tikzpicture}
  \vspace{-0.6cm}
  \caption{Speedup comparison for eight custom instructions generated under different \gls{io} constraints.}
  \label{fig:io_plot}
  \vspace{-0.3cm}
\end{figure}

\section{Results \& Discussion}
\label{section:results-and-discussion}

\begin{table}[b!]
  \vspace{-0.3cm}
  \centering
  \caption{Evaluation results. For each benchmark and \gls{io} configuration, following data is shown: Total number of valid subgraphs ($|S|$) and isomorphism classes ($|U|$), cycle based speedup ($\frac{f_{b}}{f_{c}}$), clock period increase over the baseline processor ($T_{\text{clk}}\uparrow$), and execution time speedup ($\frac{T_{b}}{T_{c}}$). Eight instructions are generated for each benchmark and \gls{io} configuration.}
  \label{tab:results}
  \begin{tabular}{lcccccc}
    \toprule
    \textbf{Benchmark} & \textbf{\gls{io}} & $\bm{|S|}$ & $\bm{|U|}$ & $\bm{\frac{f_{b}}{f_{c}}}$ & \textbf{$T_{\text{clk}}\uparrow$} & $\bm{\frac{T_{b}}{T_{c}}}$ \\
    \midrule
    & 2, 1 & 432 & 324 & 1.20 & 1.2\% & 1.18 \\
    mb-bitcnts & 3, 1 & 678 & 438 & 1.22 & 0.0\% & 1.22 \\
     & 3, 2 & 5556 & 3254 & 1.48 & 0.6\% & 1.47 \\
    \midrule
     & 2, 1 & 409 & 289 & 1.00 & 1.2\% & \textbf{0.99} \\
    mb-susan & 3, 1 & 656 & 385 & 1.19 & 1.2\% & 1.17 \\
     & 3, 2 & 5492 & 2954 & 1.49 & 0.6\% & 1.48 \\
    \midrule
     & 2, 1 & 268 & 190 & 1.55 & 2.4\% & 1.52 \\
    eb-aha-mont64 & 3, 1 & 476 & 266 & 1.60 & 2.4\% & 1.56 \\
     & 3, 2 & 2893 & 1744 & 2.14 & 2.4\% & 2.09 \\
    \midrule
    & 2, 1 & 625 & 269 & 1.79 & 1.2\% & 1.77 \\
    eb-aes & 3, 1 & 939 & 350 & 1.98 & 0.0\% & 1.98 \\
     & 3, 2 & 65894 & 4802 & 2.50 & 1.2\% & \textbf{2.47} \\
    \midrule
     & 2, 1 & 231 & 188 & 1.44 & 3.0\% & 1.39 \\
    eb-crc32 & 3, 1 & 359 & 254 & 1.64 & 3.0\% & 1.59 \\
     & 3, 2 & 2257 & 1638 & 1.91 & 0.0\% & 1.91 \\
    \midrule
     & 2, 1 & 241 & 194 & 1.05 & 0.6\% & 1.04 \\
    eb-md5sum & 3, 1 & 384 & 266 & 1.27 & 1.2\% & 1.25 \\
     & 3, 2 & 2719 & 1872 & 1.54 & 3.0\% & 1.49 \\
    \midrule
     & 2, 1 & 1971 & 364 & 1.87 & 0.0\% & 1.87 \\
    eb-sha256 & 3, 1 & 6775 & 852 & 2.29 & 0.0\% & 2.29 \\
     & 3, 2 & 790357 & 11455 & 2.37 & 0.0\% & 2.37 \\
    \midrule
     & 2, 1 & 228 & 185 & 1.19 & 0.6\% & 1.18 \\
    eb-tarfind & 3, 1 & 355 & 251 & 1.24 & 4.9\% & 1.18 \\
     & 3, 2 & 2282 & 1645 & 1.36 & 3.7\% & 1.32 \\

\bottomrule
\end{tabular}
\end{table}

The following section presents the evaluation setup and the results of the custom instruction generation using \gls{cid} for a set of software benchmarks.

For the evaluation of our framework, we selected eight applications: two from the MiBench (\textit{mb}) and six from the Embench (\textit{eb}) benchmark suites.
These benchmarks provide a set of widely used embedded applications, covering a diverse range of tasks from cryptographic algorithms to image processing.
All benchmarks are implemented in C and compiled using the RISC\nobreakdash-V GCC toolchain~\cite{riscv-gnu-toolchain}.
The code was compiled with the \textit{-O2} optimization level, enabling aggressive optimizations for improved performance, providing a more realistic baseline for comparison.
Profiling data was gathered using an in-house instruction-accurate RISC-V simulator.
Other simulators, such as Spike~\cite{spike-sim}, can also be used with minor modifications to the framework.
The synthesis results were obtained using 32/28 nm CMOS standard cell library provided by Synopsys, which serves as an academic technology process for research purposes.
The supply voltage is set to 0.85 V, and the library is characterized for a typical process corner, operating at a temperature of 25~°C.

We evaluate the tool by generating eight custom instructions for each of the benchmarks using different \gls{io} constraints. The results of the evaluation are presented in Table~\ref{tab:results}.
As the number of inputs and outputs allowed for custom instructions increases, so does the potential for performance improvement. This can be seen in the numbers of valid subgraphs and isomorphism classes.
The growth is particularly pronounced when increasing the maximum number of outputs to two as it implicitly allows disconnected subgraphs to be generated.
Another observation is that the ratio of valid subgraphs to isomorphism classes varies between benchmarks, reflecting different levels of instruction diversity. Some benchmarks, such as \textit{eb-aes} and \textit{eb-sha256}, exhibit a high recurrence of instructions, while others, like \textit{eb-md5sum} and \textit{eb-tarfind}, contain more isomorphism classes relative to the total number of subgraphs, indicating greater pattern diversity.

\begin{figure*}[t]
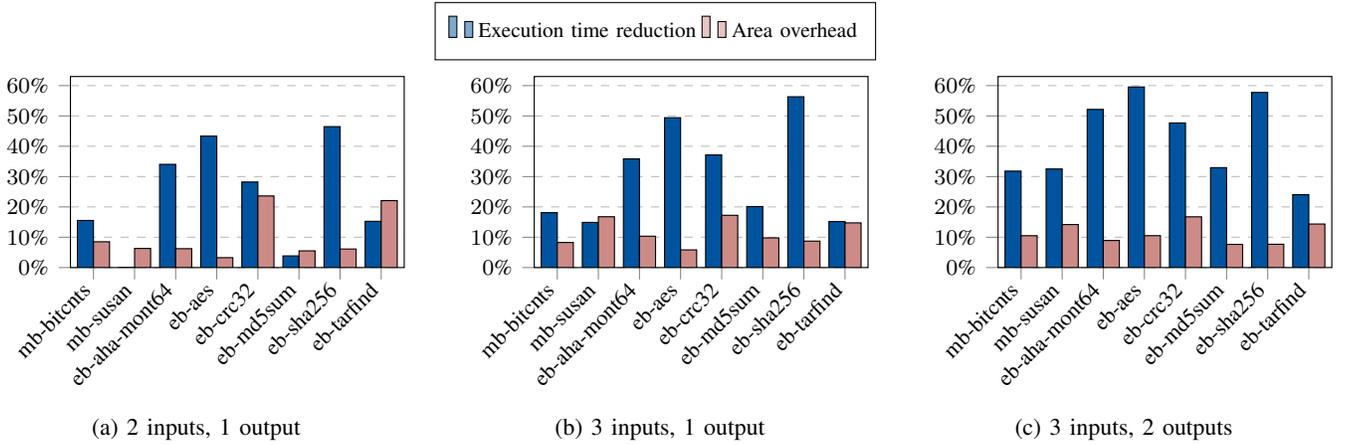

  \begin{center}
    \ref*{arealegend}
    \vspace{-0.1cm}
    \end{center}
  \centering
  \begin{subfigure}{0.32\textwidth}
    \resizebox{\textwidth}{!}{\plotarea{tikz/data/area_2_1.csv}}
    \caption{2 inputs, 1 output}
  \end{subfigure} \hfill
  \begin{subfigure}{0.32\textwidth}
    \resizebox{\textwidth}{!}{\plotarea{tikz/data/area_3_1.csv}}
    \caption{3 inputs, 1 output}
  \end{subfigure} \hfill
  \begin{subfigure}{0.32\textwidth}
    \resizebox{\textwidth}{!}{\plotarea{tikz/data/area_3_2.csv}}
    \caption{3 inputs, 2 outputs}
  \end{subfigure}
  \caption{Execution time reduction and area overhead for eight custom instructions generated for different benchmarks.}
  \label{fig:area}
  \vspace{-0.3cm}
\end{figure*}

Moreover, Table~\ref{tab:results} illustrates the impact of custom instructions on the processor's critical path, emphasizing the importance of microarchitecture awareness during selection. It compares the cycle-based speedup achieved with the selected instructions to the speedup in estimated execution time. The latter is always lower or equal to the cycle-based speedup due to the increased clock period of the customized processor. The selection process described in the previous section mitigates the selection of instructions that significantly extend the critical path. As a result, most benchmarks exhibit a moderate clock period increase, ranging from 0 to 2.4\%. However, in some cases, this value reaches up to 4.9\% making a noticeable impact on the overall performance improvement.

\newpage
Fig.~\ref{fig:io_plot} provides a graphical representation of the execution time-based performance improvement achieved by the generated custom instructions for each benchmark and \gls{io} configuration.
As expected, performance improvement increases with the number of input and output operands allowed for custom instructions, as this expands the customization potential. However, the correlation between \gls{io} constraints and performance improvement varies across benchmarks. Some exhibit a steady increase in speedup, while others follow a less predictable pattern.
The highest speedup of 2.47 is observed in the \textit{eb-aes} benchmark for the least restrictive configuration of up to three inputs and two outputs per custom instruction.

Notably, Fig.~\ref{fig:io_plot} shows one example of customization that did not lead to a performance improvement: the \textit{mb-susan} benchmark with an \gls{io} configuration of two inputs and one output. As shown in Table~\ref{tab:results}, the selected instructions offer a negligible cycle reduction (less than 1.01x speedup), which is completely outweighed by the increased clock period of the customized processor. This example underscores the importance of considering the overall processor performance impact when selecting custom instructions.

An important advantage of \gls{cid} over existing approaches is its ability to provide accurate area overhead estimates for the generated custom instructions.
Area overhead typically reflects the cost of performance gains achieved by implementing additional instructions in hardware.
Therefore, Fig.~\ref{fig:area} compares area overhead with performance improvement.
Performance improvement is measured as the percentage reduction in the estimated execution time of the application on the customized \gls{asip} relative to the baseline processor.
Area overhead is defined as the increase in processor area due to the custom \gls{fu}, compared to the baseline processor's area, as reported by the synthesis framework.
While the relative importance of these values depends on specific design requirements, this comparison provides a comprehensive view of the trade-off between performance and area cost.

Across all benchmarks and \gls{io} configurations, the area increase remains below 24\%, with an average of 11.3\%.
The area overhead varies significantly across benchmarks, reflecting differences in the selected custom instructions. Benchmarks dominated by bitwise operations, such as \textit{eb-sha256} and \textit{eb\nobreakdash-aes}, exhibit the lowest area overhead and thus the best trade-off between performance improvement and area cost.
In contrast, benchmarks such as \textit{eb-crc32} or \textit{mb-susan} show higher area overhead due to the inclusion of multiplication among the selected instructions.
Unlike performance improvement, area overhead does not directly correlate with the number of inputs and outputs for custom instructions. Instead, it is mainly influenced by the complexity of the selected instructions rather than the number of operands.
\section{Conclusion \& Outlook}
\label{section:conclusion-and-outlook}

This work focused on advancing automated processor customization within the RISC\nobreakdash-V ecosystem.
While identifying custom instructions in applications is thoroughly studied field of research,
accurately assessing the merit of the proposed candidates is a weak spot of most works as they rely on overly simplified models.

We address this issue by proposing \gls{cid}, a framework that generates candidates in the nML processor modeling language.
These descriptions can be integrated into tools such as Synopsys ASIP Designer, where the instructions are incorporated into a baseline processor.
The resulting processor can then be synthesized, enabling unparalleled degrees of accuracy in estimating both performance and area.

We evaluated our methodology through several case studies using various benchmarks and configurations.
The results showed significant performance gains across different workloads, with speedups of up to 2.47x compared to a baseline RISC\nobreakdash-V processor.
The attained speedups came at the expense of less than 24\% in area overhead.

Building on our promising results, several opportunities for future exploration remain.
One direction is to extend the range of supported customizations, for instance, by adding floating-point operations, memory accesses, or vector instructions.
Additionally, integrating compiler support for custom instructions presents an opportunity to create a more seamless end-to-end solution, allowing the compiler to efficiently recognize and leverage the instructions generated by our framework.

\newpage
\bibliographystyle{IEEEtran}
\balance
\bibliography{refs}

\end{document}